\documentclass[journal=ancac3,manuscript=article,email=true]{achemso}

\usepackage[version=3]{mhchem} % Formula subscripts using \ce{}
\usepackage{upgreek}

\author{Rieke von Seggern}
\affiliation[Physics UR]{Department of Physics and Regensburg Center for Ultrafast Nanoscopy (RUN), University of Regensburg, Germany}
\author{Jasmin Pongratz}
\affiliation[Physics UR]{Department of Physics and Regensburg Center for Ultrafast Nanoscopy (RUN), University of Regensburg, Germany}
\author{Christine Ziegler}
\affiliation[Biology UR]{Department of Biophysics II/Structural Biology and Regensburg Center for Ultrafast Nanoscopy (RUN), University of Regensburg, Germany}
\author{Sascha Schäfer}
\affiliation[Physics UR]{Department of Physics and Regensburg Center for Ultrafast Nanoscopy (RUN), University of Regensburg, Germany}
\email{sascha.schaefer@ur.de}

\title{Visualizing Transient Ordering Phenomena in Dense Nanoparticle Clouds}

\keywords{Liquid-phase transmission electron microscopy, nanofluidics, gold nanoparticles, diffusion, nanoscale confinement, particle-particle interaction potential}

\begin{document}

\begin{abstract}
The dynamics of nanoparticles within nanoscale liquid environments exhibit a range of complex phenomena driven by the interplay of processes at varying length scales. While these dynamics have profound technical implications, such as in nanoscale catalytic kinetics, ion-transport pathways in energy storage, and macromolecular crowding in biological systems, real-space imaging of dense, confined nanoparticle assemblies remains a significant challenge.  
Here, we present a liquid-phase transmission electron microscopy approach in which dense clouds of gold nanoparticles are formed within microfluidic channels, rendering the particle ensemble visible in bright-field electron imaging. This strategy enables direct imaging of different density-dependent particle ordering phenomena, including a local structuring of the colloidal liquid in nanoscale spaces, disordered dynamic clouds at high nanoparticle densities and the reversible formation of superlattice structures. Our results provide a unique window into the complex processes of colloidal self-organization at the nanoscale.
\end{abstract}

\section{Introduction}
Liquids in confined nanoscale spaces show properties remarkably different from their corresponding bulk phases\cite{granickMotionsRelaxationsConfined1991,schochTransportPhenomenaNanofluidics2008}, as manifested for example in anomalous fluid dynamics and diffusion\cite{rossiEnvironmentalScanningElectron2004,goertzHydrophilicityViscosityInterfacial2007,ortiz-youngInterplayApparentViscosity2013,zhongExploringAnomalousFluid2020}, ion-selective transport\cite{doyleStructurePotassiumChannel1998}, and the emergence of novel liquid phases with complex particle ordering \cite{baiPolymorphismPolyamorphismBilayer2012,jineshExperimentalEvidenceIce2008,giovambattistaEffectPressurePhase2006}. These peculiar behaviors originate from the increased importance of surface-bulk interactions, the breaking of translational symmetry by the confinement and, at the smallest length scales, by additional quantum effects. Nanofluidics has found applications in such diverse fields as materials science\cite{xuNanofluidicsNewArena2018}, energy research\cite{xuAdvancedNanoscaleFunctionalities2025}, and life sciences\cite{yangTranslocationDNAUltrathin2021} by providing critical insights into nucleation and growth phenomena\cite{zhengObservationSingleColloidal2009}, the dynamic transformation of functional materials within batteries and fuel cells\cite{yuanApplicationSituLiquid2023,soleymaniChallengesOpportunitiesUnderstanding2022, chenRationalElectrocatalystDesign2025}, and the spatial control of biomolecular processes\cite{iarossiEmergenceNanofluidicsSingleBiomolecule2025}. 

One important aspect in nanofluidics is the investigation of transport scalings for atomic- and nanoscale particles within confined spaces, including ions, colloidal particles, molecules and proteins. To experimentally access these complex and often transient phenomena in nanoscale liquids, a wide range of optical and X-ray-based approaches has been developed, each addressing different aspects of nanoscale behavior. Interferometric scattering microscopy \cite{ginsbergInterferometricScatteringMicroscopy2025,lindforsDetectionSpectroscopyGold2004,arbouetDirectMeasurementSingleMetalCluster2004}, X-ray scattering methods \cite{inghamXrayScatteringCharacterisation2015}, and single-particle fluorescence techniques \cite{mocklSuperresolutionMicroscopySingle2020,chenOpticalSuperResolutionImaging2017,thompsonPreciseNanometerLocalization2002,pertsinidisSubnanometreSinglemoleculeLocalization2010} have enabled the mapping of structural dynamics, particle aggregation, and conformational changes in real time. These approaches can provide single-particle sensitivity and fast temporal resolution, but are often limited in the achievable spatial resolution, require ensemble averaging in dense systems, and provide only indirect structural information.

To overcome some of these limitations, liquid-phase transmission electron microscopy (LPTEM) has emerged as a powerful real-space imaging methodology capable of directly visualizing nanoscale objects and their dynamics in liquid environments\cite{rossOpportunitiesChallengesLiquid2015}. Despite technical difficulties \cite{dejongeResolutionAberrationCorrection2019} and challenges originating from unavoidable electron-beam-induced radiolysis \cite{schneiderElectronWaterInteractions2014,groganBubblePatternFormation2014,fritschInfluenceIonizingRadiation2025a}, LPTEM has been applied to study a variety of systems, including the dynamics and structure of biomolecules \cite{jongeElectronMicroscopyWhole2009, mohantyImpermeableGraphenicEncasement2011, chenStudiesDynamicsBiological2014, smithElectronVideographyLipid2024}, nanoparticle growth \cite{zhengObservationSingleColloidal2009, liDirectionSpecificInteractionsControl2012, liaoRealTimeImagingPt3Fe2012a, lohMultistepNucleationNanocrystals2017} and self-assembly \cite{parkDirectObservationNanoparticle2012a, ouKineticPathwaysCrystallization2020, cepeda-perezElectronMicroscopyNanoparticle2020a, arenasestebanQuantitative3DStructural2024}, that are difficult to access with purely optical techniques.

The dynamics of metal nanoparticles (NPs)\cite{woehlMechanismsNanoparticleSurface2015a,parentTacklingChallengesDynamic2018a} are influenced by many factors, including Brownian motion, hydrodynamic interactions, surface forces, and collective effects that emerge at high particle concentrations. LPTEM experiments have enabled the direct observation of particle motion in liquid, monitoring their translational and rotational dynamics as they approach each other, which yields information on interparticle forces\cite{liaoRealTimeImagingPt3Fe2012a, zhengElectronBeamManipulation2012, liDirectionSpecificInteractionsControl2012, kangRealspaceImagingNanoparticle2021a}. The movement of NPs in ultrathin liquid layers is slowed down by multiple orders of magnitude\cite{groganSituLiquidcellElectron2011, ringVideofrequencyScanningTransmission2012, whiteChargedNanoparticleDynamics2012a, luNanoparticleDynamicsNanodroplet2014, woehlMechanismsNanoparticleSurface2015a, chenSituWetcellTEM2012, liuSituVisualizationSelfAssembly2013, verchExceptionallySlowMovement2015a, cheeDirectObservationsRotation2019,cheeDesorptionMediatedMotionNanoparticles2016a, cazadeStructureDynamicsElectrolyte2014,bakalisComplexNanoparticleDiffusional2020,parentTacklingChallengesDynamic2018a}, while in less-confined liquid-cells NPs show Brownian motion with diffusion coefficients comparable to those predicted by the Stokes–Einstein relation\cite{yesibolatiUnhinderedBrownianMotion2020a,jamaliAnomalousNanoparticleSurface2021a}.

Apart from single-particle dynamics, LPTEM studies have also provided insight into NP self-assembly, driven by external convection\cite{parkDirectObservationNanoparticle2012a} or NP--NP interactions\cite{liuSituVisualizationSelfAssembly2013,powersTrackingNanoparticleDiffusion2017,sutterSituMicroscopySelfassembly2016,chenInteractionPotentialsAnisotropic2015,welchUnderstandingRoleSolvation2016,anandHydrationLayerMediatedPairwise2016,luoQuantifyingSelfAssemblyBehavior2017,liuColloidAtomDualityAssembly2020}. Characteristics of nonclassical crystallization, such as the coalescence of assembled NPs and the formation of disordered aggregates prior to periodic self-assembly, were observed during the self-assembly process.\cite{parkDirectObservationNanoparticle2012a, powersTrackingNanoparticleDiffusion2017, chenNucleationGrowthSuperlattice2020}
Recently, small-angle X-Ray scattering demonstrated the formation of a dense metastable colloidal fluids before a transformation to a quasi-crystalline phase of agglomerated NPs occurred \cite{coropceanuSelfassemblyNanocrystalsStrongly2022, tannerEnhancingNanoscaleCharged2025a}. Real-space imaging experiments on such dense nanoparticle phases are so-far missing.

Here, we image ordering phenomena of gold nanoparticles in aqueous solution by bright-field transmission electron microscopy. After introducing a dilute solution of citrate-capped gold NPs into microfluidic channels embedded in an electron-transparent membrane, we locally raise the particle density in situ by immobilizing particles into a sieve-like blockage that retains incoming NPs while passing the solvent. This approach enables direct, real-space observation of density-dependent ordering regimes that have so far been accessible only through scattering techniques, including a confinement-induced structuring of the colloidal liquid that correlates with the sites of subsequent particle immobilization, disordered, dynamic clouds reminiscent of liquid–liquid phase separation, and the reversible, flow-dependent formation of nanoparticle superlattices.

\section{Experiments and discussion}
\begin{figure}
    \includegraphics[width=\linewidth]{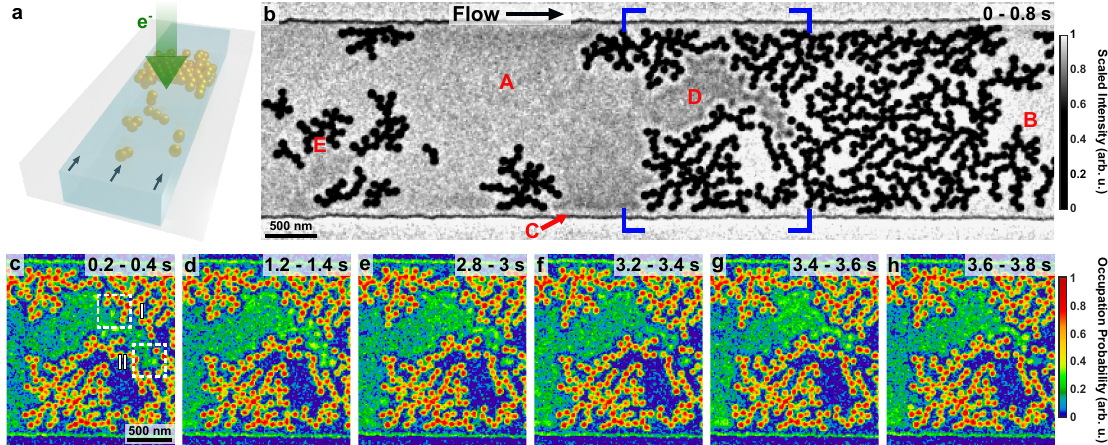}
    \caption{Visualizing the NP density in a flowing colloid near a nano-constriction. (a) Experimental scheme of colloidal solution flowing through a microchannel embedded in a silicon nitride membrane. (b) Bright-field electron micrograph of an \textit{in situ} deposited nano-constriction formed by deposited NPs. Differences in the image contrast of the flowing colloid indicate the variation of the local NP density. Contrast in A and B corresponds to solution with and without NPs. Close to the channel edge (label C)  and in funnel-like structures (D+E) the NP density is locally enhanced. (c-h) Evolution of the particle distribution close to the constriction over a 4 s duration, demonstrating the preferred localization of colloidal NPs within the tight constriction boundaries and the correlation of these local density maxima with the final positions at which they are deposited (highlighted in areas I and II). Electron dose rate: 0.03~e$^-$\AA{}$^{-2}$s$^{-1}$. Channel cross-section: 180-210~nm $\times$ 2~µm.}
    \label{fig:1}
\end{figure}

For imaging colloidal ordering phenomena on nanometer length scales in TEM, we focus on the behavior of citrate-capped gold NPs (60~nm diameter, see Methods) in aqueous solution as a simple model system. The colloidal solution is injected into the microchannels of an electron-beam-transparent liquid-cell chip allowing for \textit{in situ} nanoscale TEM imaging of the flowing solution. The fluid flow can be controlled by the externally applied pressure at the chip inlet with additional contributions from capillary forces present for  partially filled chips. 

In the freely flowing solution, individual nanoparticles cannot be resolved. At exposure times of about 100~ms used here, Brownian motion smears each particle across the image, leaving no discernible single-particle contrast. To establish a fixed starting point, we therefore immobilize particles at a chosen location. Focusing the electron beam onto a selected position in the microfluidic channel raises the local dose rate and pins nanoparticles against the channel walls. Due to the continuous influx of colloidal solution through the microchannels, we can locally deposit densely packed dendritic NP agglomerates, as shown in Fig.~1(b), reaching particle area densities of 80~µm$^{-2}$ (volume density of 400~µm$^{-3}$) far exceeding the NP density in solution of 0.46 $\upmu$m$^{-2}$. The immobilized particles act as a sieve, so that only the solvent passes through this blockage. The corresponding difference in local NP density in solution is clearly visible in the change of image brightness in liquid-filled regions up- and downstream from the blockage (comparing regions A and B in Fig.~1(b)). 

Interestingly, also the contrast upstream of the blockages markedly varies between different regions both in amplitude and in homogeneity. Despite fast Brownian motion, in certain image areas an increased particle density can be found, such as at channel borders (point C) or in funnel-like structures (points D, E). Furthermore, in the tip of the funnel D,  the smooth particle cloud breaks up into a grainy density structure, in which individual NPs already seem to become discernible, indicating that translational symmetry in the colloidal liquid is broken. Remarkably, the contrast at each of these positions is only about 80\% of the contrast of immobilized particle, demonstrating that this ordering phenomenon occurs at particle occupation probabilities smaller than one and entails significant fluctuations. The confinement-induced spatial structure of the colloid is stable for a prolonged time, provided that the confinement and flow conditions persist. For example for an image recorded 1.2 s later (Fig.~1(d)), the corrugation of the image contrast in the funnel essentially stays constant. The temporal evolution of the image contrast in the funnel, as shown in Fig.~1(c)–(h), further illustrates the tight connection between the funnel shape, colloidal structuring and dendritic growth. For example, for the feature in area I highlighted in Fig.~1(c) (dashed box), for more than 3 s a faint maximum in the local NP occupation in the colloid exists. Between 3.2-3.4~s the maximum sharpens until in the next frame, the image contrast is equal to other immobilized particles. This highlights that sites with a high occupation probability in the liquid  correlate with the position at which particles get subsequently immobilized on the surface (see Video 1), indicating that the local structure of the colloid patterns the growth of the NP agglomerate. During the immobilization the particle (contrast maximum) moves closer to the already existing agglomerate and in Fig.~1(g) a dark contrast around the newly immobilized particle develops similar to the excluded volume with a thickness of about 45-60~nm which is found around the whole aggregate. Such a region signifies an effective repulsive interaction between suspended and immobilized NPs which is also consistent with particle distances in the structured colloid (Fig.~1(d)) on the order of 130~nm. We hypothesize that this rather long interaction range (as compared to the particle and ligand size) is mostly governed by the ligand-induced negative particle charges in combination with screening by ions in the liquid. Once the particle comes in direct contact at the immobilized agglomerate, potentially due to a radiolytically driven\cite{fritschInfluenceIonizingRadiation2025a} gradual loss of citrate capping, the system acts as a single charged conducting surface resulting in negligible particle distances within the agglomerate. 

The evolution of the structuring within the colloid and its dependence on a changing funnel shape becomes apparent in area II (Fig.~1(c), second dashed box). Here, particles in a small region are separated from the rest through immobilization of another particle, visible in Fig.~1(e). In the closed-off area, the increased contrast in the central part of the area stays rather blurred but with a slightly higher occupation probability on the right side of the accessible space. The blurred cloud sharpens after about 0.8~s (Fig.~1(g),(h)), revealing that this cloud was actually formed by a single mobile NP.

\begin{figure}
    \includegraphics[width=\linewidth]{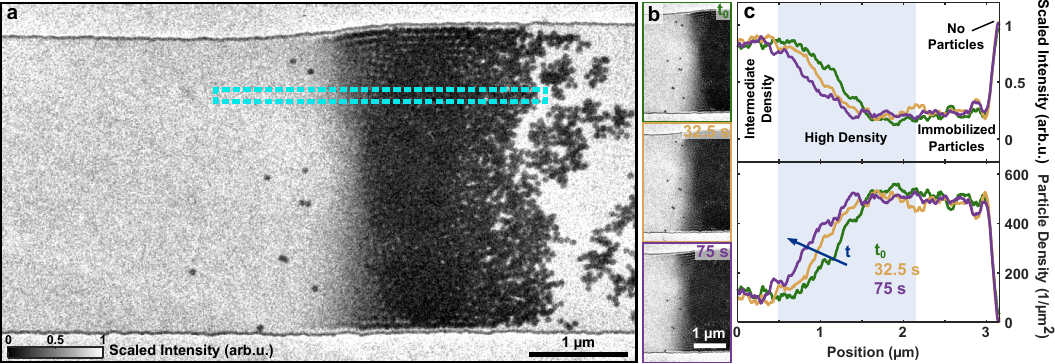}
    \caption{Formation of high-density nanoparticle cloud in front of a nano-constriction. (a) Bright-field electron micrograph of a microfluidic channel, recorded about 1 min after the NP blockage was formed. NPs in the flowing colloidal solution cannot pass through the constriction and accumulate in front of the structure in a dense particle cloud. (b) Evolution of the boundary between the high-density NP cloud and the streaming colloid front over 75~s duration, indicating a front velocity of 4.9~nm/s. (c) Image intensity (top panel) in area enclosed by broken line in (a) at three different times (laterally averaged over width as indicated). NP area density (bottom panel) obtained from the image intensity. See text for details. Electron dose rate: 0.02~e$^-$\AA{}$^{-2}$s$^{-1}$. Channel cross-section: 310~nm $\times$ 3~µm.}
    \label{fig:2}
\end{figure}

To further investigate NP--NP interaction in the colloid, we adopt experimental conditions for which the arriving NPs are only slowly deposited at the blockage site, leaving a high-density cloud of suspended NPs in front of it. An example for this scenario is displayed in Fig.~2(a). Because of the larger channel height ($\sim$310~nm) used here compared with the experiment of Fig.~1, NPs overlap in the projection image and can no longer be resolved individually. Over time, the dark contrast region shifts further upstream, as shown in Fig.~2(b), reflecting the growing volume of the high-density cloud as new NPs continue to arrive. Line profiles (Fig.~2(c)) show that the particle density, which varies inversely with the image brightness, increases towards the blockage  from the direction of the incoming fluid. 

To quantify this, we assume that the image brightness decreases linearly with NP coverage, calibrated so that NP-free solvent (brightness $I_{max}$) corresponds to zero density and a single particle of cross-section $A_{NP}=\pi r^2$ (brightness $I_{single}$) to a density of $1/A_{NP}$. As no suspended NPs are present downstream of the blockage, the local area density follows as $n=\frac{1}{A_{NP}}\left(1-(I-I_{single})/(I_{max}-I_{single})\right)$, with $I$ the local pixel brightness. The resulting densities (Fig.~2(c), bottom) exceed that of the input solution (0.71~µm$^{-2}$) by almost three orders of magnitude and reveal a clear separation between an intermediate-density phase ($\sim$120~µm$^{-2}$, left of Fig.~2(a)) and a high-density phase ($\sim$500~µm$^{-2}$). This apparent liquid--liquid phase separation in a highly concentrated colloid echoes early light-scattering experiments on sub-micrometer polymer spheres\cite{puseyPhaseBehaviourConcentrated1986} and recent small-angle X-ray scattering on few-nanometer semiconductor NPs\cite{tannerEnhancingNanoscaleCharged2025a}, but here resolves the structural evolution in real space. For example, a series of electron micrographs recorded over 75 s (see Video 2) highlights the fluctuating behavior of the boundary and even whirl-like features in the intermediate-density phase.  

We note that from the gradual upstream shift of the phase boundary together with the NP density of the phases, we can also infer a local liquid flow velocity on the order of 3.4~µm/s, assuming that the required number of NPs is supplied by an influx at the density of the initial solution (see Supplement).

\begin{figure}
    \centering
    \includegraphics[width=\linewidth]{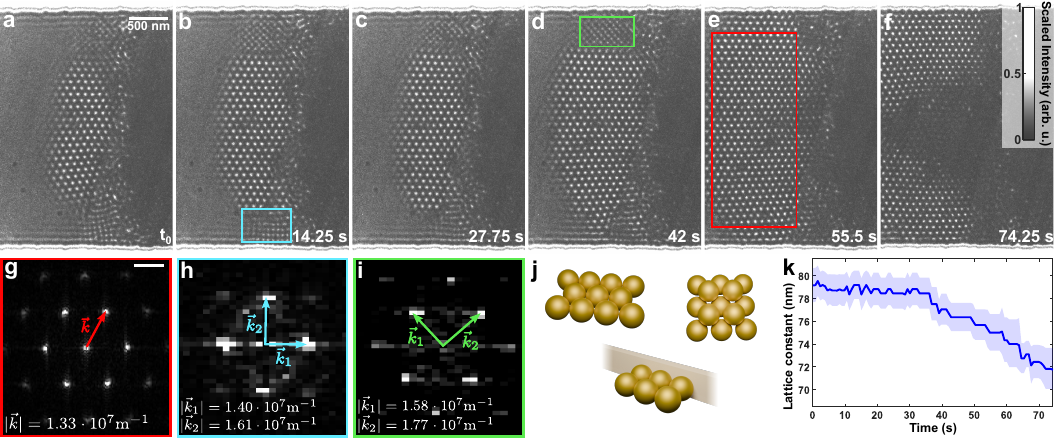}
    \caption{Emergence of superstructure in high-density particle cloud. (a-f) Electron micrographs of a cloud section close to the channel blockage several minutes after forming the constriction. Particles in the cloud start to form transient lattice structures in solution which disappear when removing the solvent flow. In the central part of the channel, the superlattice shows hexagonal symmetry whereas at the channel edges a rectangular lattice is found. Both lattices are roughly aligned with the channel orientation. (g-i) Spatial Fourier transforms of the image intensity within the colored rectangles indicated in (e,b,d), respectively. (j) Structural models for hexagonal close-packed and quadratic NP arrangements, respectively. (k) Evolution of the lattice constant in the hexagonal central domain, indicating a compactification putatively related to a ligand loss and particle decharging. Electron dose rate: 0.1~e$^-$\AA{}$^{-2}$s$^{-1}$. Channel cross-section: 310~nm $\times$ 3~µm.}
    \label{fig:3}
\end{figure}

Within the high-density colloid phase, we find densities approaching those expected for a close-packed NP crystal. For five (111) layers of a face-centered cubic lattice of solid spheres of 60~nm diameter, a value of about 1600~$\upmu$m$^{-2}$ is expected (see Supplement). At such densities, pronounced NP--NP interactions become relevant, and indeed we observe the emergence of an ordered superstructure of mobile NPs. This spatial ordering in solution is illustrated by the frame sequence in Fig.~3(a)--(f) (see also Video~3). Shortly after ordering sets in (Fig.~3(a)), NPs across much of the dense cloud form a lattice with hexagonal symmetry. The channel edges initially hinder hexagonal ordering (Fig.~3(b),(d)), producing domains with a cubic-like arrangement. Over time, the hexagonal domain grows and eventually extends into the edge regions. Turning off the flow breaks up the lattice (see Video~4), highlighting the transient nature of the superlattice order and confirming that it is formed by suspended (non-immobilized) particles.

A spatial Fourier transform of the hexagonal domain contrast (taken from red area in Fig.~3(e)) yields a lattice constant of about $1/k=75$~nm, notably larger than the 60~nm NP diameter. This corresponds to a surface-to-surface separation of only about 15~nm, compared with the 45~nm particle-free region around each immobilized particle in Fig.~1. Several aspects may affect the potential minima in the two cases. First, the excluded region in Fig.~1 reflects the interaction between immobilized and free particles, whereas the lattice reflects the interaction between free particles only. However, immobilized particles are expected to carry fewer negatively charged citrate ligands than free particles, so that a less-repulsive interaction is expected, contrary to the experimental findings. Second, the effective interaction depends strongly on charge screening, so that the local ion concentration is an important factor in setting the minimum particle distance. Third, the streaming solution exerts a pressure that further compresses the lattice. As the superlattice dissolves once the flow is stopped, this compression appears to be a necessary condition for lattice formation under our experimental conditions.

Extracting the hexagonal lattice constant from the Fourier transform of each frame (Fig.~3(k)) shows that the interparticle distance decreases over time, approaching the nominal particle diameter. Although the local ion concentration is expected to depend on the electron dose rate\cite{fritschInfluenceIonizingRadiation2025a}, the time evolution of the lattice constant---an initial plateau followed by a linear decrease---suggests that the interparticle spacing is governed more by the growth of the superlattice. In particular, the linear decrease appears to set in once the superlattice coherently spans the entire channel width. This behavior shows that the interparticle interaction in our experiments is far from constant and calls for theoretical models that account for the local ionic strength, the evolving ligand coverage of the particles, and the strain imparted by the streaming solution.

\section{Conclusion}

In summary, we have demonstrated real-space imaging of dense gold nanoparticle clouds in solution by microfluidic liquid-phase TEM, providing direct access to colloidal ordering in dense, confined systems that have so far been studied mainly by ensemble-averaging scattering techniques. By locally increasing the particle density within microfluidic channels, we render the ensemble of suspended particles visible as a spatially varying contrast that reflects the local density, and thereby resolve several density-dependent regimes of colloidal ordering.

At low densities, the colloid develops a spatial patterning of the contrast in strongly confined regions, and particles subsequently adhere one by one to a growing dendritic agglomerate. Notably, the local pattern in the colloid correlates with the sites at which particles are subsequently immobilized. At intermediate densities, larger numbers of suspended particles form disordered, dynamic clouds reminiscent of a liquid--liquid phase separation. At the highest densities, the particles reversibly assemble into superlattices. All of these regimes are expected to be governed by the interparticle interaction potential, which itself depends on the local environment---the confinement geometry, the ion concentration, and the proximity to the channel surfaces---and, in our experiments, also on the electron irradiation and the imposed flow.

Finally, the interplay between confinement, density, and interaction potential observed here for nanoparticles may offer a conceptual model for molecular and ionic transport in atomically tight channels, such as those found in membrane proteins. While the relevant length scales differ by orders of magnitude, the shared physics of crowding and screened interactions under confinement suggests that real-space electron imaging of nanoparticle model systems could complement studies of transport at the molecular scale.

\section{Methods}
For the liquid-phase experiments, a sample holder and corresponding chips by Insight Chips have been used together with a flow control unit and accessories by Fluigent (see Supplement Fig.~S1). Before every measurement, the holder was flushed with Millipore-filtered water and dried with air to minimize the possibility of built-up residues interfering with the experiment. After cleaning, a new chip was installed at the tip and sample fluid was pumped into the holder while observing the contrast of the channels in the electron-transparent silicon-nitride window by optical microscopy. Once a change in color of the visible channels indicated their successful filling, the sample holder was inserted into the TEM.

TEM measurements were conducted using a JEOL F200 TEM with an accelerating voltage of 200~kV, equipped with the XF416R camera by TVIPS. All micrographs were recorded in MAG mode at different magnifications (5k, 8k and 15k for Figures 1,2 and 3, respectively) and without inserted objective aperture. Fig.~1(b) is a sum over four frames with an exposure time of 200~ms each, while Fig.~1(c)-(h) show micrographs at the original exposure time. The data in Fig.~2 and Fig.~3 was compiled by summing over 5 frames with an exposure time of 150~ms each.

A chip with channel height of approx. 180-210~nm and channel width of 2~µm was used to record the data shown in Fig.~1. For Fig.~2 and Fig.~3 the channel height and width was approx. 310~nm and 3~µm, respectively.

Citrate-capped gold NPs with a diameter of 60~nm were purchased from Nanopartz Inc. and employed as-is in an aqueous solution at a concentration of about 2.3$\times 10^{12}$ NPs/ml (5 mg Au/ml). For a liquid film height of 200~nm, such a volume density corresponds to an area density of 0.46~µm$^{-2}$. At the electron imaging conditions in Fig.~2, a single (immobilized) gold particle shows a contrast of about 55 \%.

\begin{acknowledgement}
This work was supported by the Deutsche Forschungsgemeinschaft (DFG, German Research Foundation) through GRK 2905 – project-ID 502572516. The authors thank E. C. S. Jensen and Insight Chips for their continuous support regarding the sample holder and chips.

\end{acknowledgement}
\newpage

\begin{suppinfo}
Scheme of liquid-cell setup; Estimates on fluid velocity, fcc area density and electron dose rate; Videos of data from Fig.~1-3; Additional video of dissolving superlattice after external pressure was turned off. Videos are available under: https://epub.uni-regensburg.de/79564/.
\end{suppinfo}
\newpage

\bibliography{NPbib}

@article{anandHydrationLayerMediatedPairwise2016,
  title = {Hydration {{Layer-Mediated Pairwise Interaction}} of {{Nanoparticles}}},
  author = {Anand, Utkarsh and Lu, Jingyu and Loh, Duane and Aabdin, Zainul and Mirsaidov, Utkur},
  year = 2016,
  month = jan,
  journal = {Nano Letters},
  volume = {16},
  number = {1},
  pages = {786--790},
  publisher = {American Chemical Society},
  issn = {1530-6984},
  doi = {10.1021/acs.nanolett.5b04808},
  urldate = {2025-12-17}
}

@article{arbouetDirectMeasurementSingleMetalCluster2004,
  title = {Direct {{Measurement}} of the {{Single-Metal-Cluster Optical Absorption}}},
  author = {Arbouet, A. and Christofilos, D. and Del Fatti, N. and Vall{\'e}e, F. and Huntzinger, J. R. and Arnaud, L. and Billaud, P. and Broyer, M.},
  year = 2004,
  month = sep,
  journal = {Physical Review Letters},
  volume = {93},
  number = {12},
  pages = {127401},
  publisher = {American Physical Society},
  doi = {10.1103/PhysRevLett.93.127401},
  urldate = {2025-12-17}
}

@article{arenasestebanQuantitative3DStructural2024,
  title = {Quantitative {{3D}} Structural Analysis of Small Colloidal Assemblies under Native Conditions by Liquid-Cell Fast Electron Tomography},
  author = {Arenas Esteban, Daniel and Wang, Da and Kadu, Ajinkya and Olluyn, Noa and {S{\'a}nchez-Iglesias}, Ana and {Gomez-Perez}, Alejandro and {Gonz{\'a}lez-Casablanca}, Jes{\'u}s and Nicolopoulos, Stavros and {Liz-Marz{\'a}n}, Luis M. and Bals, Sara},
  year = 2024,
  month = jul,
  journal = {Nature Communications},
  volume = {15},
  number = {1},
  pages = {6399},
  publisher = {Nature Publishing Group},
  issn = {2041-1723},
  doi = {10.1038/s41467-024-50652-y},
  urldate = {2026-04-13},
  copyright = {2024 The Author(s)},
  langid = {english}
}

@article{baiPolymorphismPolyamorphismBilayer2012,
  title = {Polymorphism and Polyamorphism in Bilayer Water Confined to Slit Nanopore under High Pressure},
  author = {Bai, Jaeil and Zeng, Xiao Cheng},
  year = 2012,
  month = dec,
  journal = {Proceedings of the National Academy of Sciences},
  volume = {109},
  number = {52},
  pages = {21240--21245},
  publisher = {Proceedings of the National Academy of Sciences},
  doi = {10.1073/pnas.1213342110},
  urldate = {2026-01-21}
}

@article{bakalisComplexNanoparticleDiffusional2020,
  title = {Complex {{Nanoparticle Diffusional Motion}} in {{Liquid-Cell Transmission Electron Microscopy}}},
  author = {Bakalis, Evangelos and Parent, Lucas R. and Vratsanos, Maria and Park, Chiwoo and Gianneschi, Nathan C. and Zerbetto, Francesco},
  year = 2020,
  month = jul,
  journal = {The Journal of Physical Chemistry C},
  volume = {124},
  number = {27},
  pages = {14881--14890},
  publisher = {American Chemical Society},
  issn = {1932-7447},
  doi = {10.1021/acs.jpcc.0c03203},
  urldate = {2025-12-17}
}

@article{cazadeStructureDynamicsElectrolyte2014,
  title = {Structure and {{Dynamics}} of an {{Electrolyte Confined}} in {{Charged Nanopores}}},
  author = {Cazade, Pierre-Andre and Hartkamp, Remco and Coasne, Benoit},
  year = 2014,
  month = mar,
  journal = {The Journal of Physical Chemistry C},
  volume = {118},
  number = {10},
  pages = {5061--5072},
  publisher = {American Chemical Society},
  issn = {1932-7447},
  doi = {10.1021/jp4098638},
  urldate = {2025-12-17}
}

@article{cepeda-perezElectronMicroscopyNanoparticle2020a,
  title = {Electron Microscopy of Nanoparticle Superlattice Formation at a Solid-Liquid Interface in Nonpolar Liquids},
  author = {{Cepeda-Perez}, E. and Doblas, D. and Kraus, T. and {de Jonge}, N.},
  year = 2020,
  month = may,
  journal = {Science Advances},
  volume = {6},
  number = {20},
  pages = {eaba1404},
  publisher = {American Association for the Advancement of Science},
  doi = {10.1126/sciadv.aba1404},
  urldate = {2025-12-17}
}

@article{cheeDesorptionMediatedMotionNanoparticles2016a,
  title = {Desorption-{{Mediated Motion}} of {{Nanoparticles}} at the {{Liquid}}--{{Solid Interface}}},
  author = {Chee, See Wee and Baraissov, Zhaslan and Loh, N. Duane and Matsudaira, Paul T. and Mirsaidov, Utkur},
  year = 2016,
  month = sep,
  journal = {The Journal of Physical Chemistry C},
  volume = {120},
  number = {36},
  pages = {20462--20470},
  publisher = {American Chemical Society},
  issn = {1932-7447},
  doi = {10.1021/acs.jpcc.6b07983},
  urldate = {2025-12-17}
}

@article{cheeDirectObservationsRotation2019,
  title = {Direct {{Observations}} of the {{Rotation}} and {{Translation}} of {{Anisotropic Nanoparticles Adsorbed}} at a {{Liquid}}--{{Solid Interface}}},
  author = {Chee, See Wee and Anand, Utkarsh and Bisht, Geeta and Tan, Shu Fen and Mirsaidov, Utkur},
  year = 2019,
  month = may,
  journal = {Nano Letters},
  volume = {19},
  number = {5},
  pages = {2871--2878},
  publisher = {American Chemical Society},
  issn = {1530-6984},
  doi = {10.1021/acs.nanolett.8b04962},
  urldate = {2025-12-17}
}

@article{chenInteractionPotentialsAnisotropic2015,
  title = {Interaction {{Potentials}} of {{Anisotropic Nanocrystals}} from the {{Trajectory Sampling}} of {{Particle Motion}} Using in {{Situ Liquid Phase Transmission Electron Microscopy}}},
  author = {Chen, Qian and Cho, Hoduk and Manthiram, Karthish and Yoshida, Mark and Ye, Xingchen and Alivisatos, A. Paul},
  year = 2015,
  month = mar,
  journal = {ACS Central Science},
  volume = {1},
  number = {1},
  pages = {33--39},
  publisher = {American Chemical Society},
  issn = {2374-7943},
  doi = {10.1021/acscentsci.5b00001},
  urldate = {2025-12-17}
}

@article{chenNucleationGrowthSuperlattice2020,
  title = {Nucleation, Growth, and Superlattice Formation of Nanocrystals Observed in Liquid Cell Transmission Electron Microscopy},
  author = {Chen, Qian and Yuk, Jong Min and Hauwiller, Matthew R. and Park, Jungjae and Dae, Kyun Seong and Kim, Jae Sung and Alivisatos, A. Paul},
  year = 2020,
  month = sep,
  journal = {MRS Bulletin},
  volume = {45},
  number = {9},
  pages = {713--726},
  issn = {0883-7694, 1938-1425},
  doi = {10.1557/mrs.2020.229},
  urldate = {2025-12-17},
  langid = {english}
}

@article{chenOpticalSuperResolutionImaging2017,
  title = {Optical {{Super-Resolution Imaging}} of {{Surface Reactions}}},
  author = {Chen, Tao and Dong, Bin and Chen, Kuangcai and Zhao, Fei and Cheng, Xiaodong and Ma, Changbei and Lee, Seungah and Zhang, Peng and Kang, Seong Ho and Ha, Ji Won and Xu, Weilin and Fang, Ning},
  year = 2017,
  month = jun,
  journal = {Chemical Reviews},
  volume = {117},
  number = {11},
  pages = {7510--7537},
  publisher = {American Chemical Society},
  issn = {0009-2665},
  doi = {10.1021/acs.chemrev.6b00673},
  urldate = {2025-12-17}
}

@article{chenRationalElectrocatalystDesign2025,
  title = {Toward {{Rational Electrocatalyst Design}}: {{Dynamic Insights}} from {{Liquid Environmental Transmission Electron Microscopy}}},
  shorttitle = {Toward {{Rational Electrocatalyst Design}}},
  author = {Chen, Hanyang and Jiang, Ying and Wu, Hao Bin and Yuan, Wentao and Wang, Yong},
  year = 2025,
  journal = {Advanced Materials},
  volume = {37},
  number = {45},
  pages = {e06352},
  issn = {1521-4095},
  doi = {10.1002/adma.202506352},
  urldate = {2026-01-21},
  copyright = {\copyright{} 2025 Wiley-VCH GmbH},
  langid = {english}
}

@article{chenSituWetcellTEM2012,
  title = {In Situ Wet-Cell {{TEM}} Observation of Gold Nanoparticle Motion in an Aqueous Solution},
  author = {Chen, Xin and Wen, Jianguo},
  year = 2012,
  month = oct,
  journal = {Nanoscale Research Letters},
  volume = {7},
  number = {1},
  pages = {598},
  issn = {1556-276X},
  doi = {10.1186/1556-276X-7-598},
  urldate = {2025-12-17},
  langid = {english}
}

@article{chenStudiesDynamicsBiological2014,
  title = {Studies of the Dynamics of Biological Macromolecules Using {{Au}} Nanoparticle--{{DNA}} Artificial Molecules},
  author = {Chen, Qian and M.~Smith, Jessica and I.~Rasool, Haider and Zettl, Alex and Paul~Alivisatos, A.},
  year = 2014,
  journal = {Faraday Discussions},
  volume = {175},
  number = {0},
  pages = {203--214},
  publisher = {Royal Society of Chemistry},
  doi = {10.1039/C4FD00149D},
  urldate = {2025-12-17},
  langid = {english}
}

@article{coropceanuSelfassemblyNanocrystalsStrongly2022,
  title = {Self-Assembly of Nanocrystals into Strongly Electronically Coupled All-Inorganic Supercrystals},
  author = {Coropceanu, Igor and Janke, Eric M. and Portner, Joshua and Haubold, Danny and Nguyen, Trung Dac and Das, Avishek and Tanner, Christian P. N. and Utterback, James K. and Teitelbaum, Samuel W. and Hudson, Margaret H. and Sarma, Nivedina A. and Hinkle, Alex M. and Tassone, Christopher J. and Eychm{\"u}ller, Alexander and Limmer, David T. and {Olvera de la Cruz}, Monica and Ginsberg, Naomi S. and Talapin, Dmitri V.},
  year = 2022,
  month = mar,
  journal = {Science},
  volume = {375},
  number = {6587},
  pages = {1422--1426},
  publisher = {American Association for the Advancement of Science},
  doi = {10.1126/science.abm6753},
  urldate = {2025-12-17}
}

@article{dejongeResolutionAberrationCorrection2019,
  title = {Resolution and Aberration Correction in Liquid Cell Transmission Electron Microscopy},
  author = {{de Jonge}, Niels and Houben, Lothar and {Dunin-Borkowski}, Rafal E. and Ross, Frances M.},
  year = 2019,
  month = jan,
  journal = {Nature Reviews Materials},
  volume = {4},
  number = {1},
  pages = {61--78},
  publisher = {Nature Publishing Group},
  issn = {2058-8437},
  doi = {10.1038/s41578-018-0071-2},
  urldate = {2025-12-17},
  copyright = {2018 Springer Nature Limited},
  langid = {english}
}

@article{doyleStructurePotassiumChannel1998,
  title = {The {{Structure}} of the {{Potassium Channel}}: {{Molecular Basis}} of {{K}}+ {{Conduction}} and {{Selectivity}}},
  shorttitle = {The {{Structure}} of the {{Potassium Channel}}},
  author = {Doyle, Declan A. and Cabral, Jo{\~a}o Morais and Pfuetzner, Richard A. and Kuo, Anling and Gulbis, Jacqueline M. and Cohen, Steven L. and Chait, Brian T. and MacKinnon, Roderick},
  year = 1998,
  month = apr,
  journal = {Science},
  volume = {280},
  number = {5360},
  pages = {69--77},
  publisher = {American Association for the Advancement of Science},
  doi = {10.1126/science.280.5360.69},
  urldate = {2025-12-17}
}

@article{fritschInfluenceIonizingRadiation2025a,
  title = {The {{Influence}} of {{Ionizing Radiation}} on {{Quantification}} for {{In Situ}} and {{Operando Liquid-Phase Electron Microscopy}}},
  author = {Fritsch, Birk and Lee, Serin and K{\"o}rner, Andreas and Schneider, Nicholas M. and Ross, Frances M. and Hutzler, Andreas},
  year = 2025,
  journal = {Advanced Materials},
  volume = {37},
  number = {13},
  pages = {2415728},
  issn = {1521-4095},
  doi = {10.1002/adma.202415728},
  urldate = {2026-01-21},
  copyright = {\copyright{} 2025 The Author(s). Advanced Materials published by Wiley-VCH GmbH},
  langid = {english}
}

@article{ginsbergInterferometricScatteringMicroscopy2025,
  title = {Interferometric Scattering Microscopy},
  author = {Ginsberg, Naomi S. and Hsieh, Chia-Lung and Kukura, Philipp and Piliarik, Marek and Sandoghdar, Vahid},
  year = 2025,
  month = apr,
  journal = {Nature Reviews Methods Primers},
  volume = {5},
  number = {1},
  pages = {23},
  publisher = {Nature Publishing Group},
  issn = {2662-8449},
  doi = {10.1038/s43586-025-00391-1},
  urldate = {2025-12-11},
  copyright = {2025 Springer Nature Limited},
  langid = {english}
}

@article{giovambattistaEffectPressurePhase2006,
  title = {Effect of Pressure on the Phase Behavior and Structure of Water Confined between Nanoscale Hydrophobic and Hydrophilic Plates},
  author = {Giovambattista, Nicolas and Rossky, Peter J. and Debenedetti, Pablo G.},
  year = 2006,
  month = apr,
  journal = {Physical Review E},
  volume = {73},
  number = {4},
  pages = {041604},
  publisher = {American Physical Society},
  doi = {10.1103/PhysRevE.73.041604},
  urldate = {2026-01-21}
}

@article{goertzHydrophilicityViscosityInterfacial2007,
  title = {Hydrophilicity and the {{Viscosity}} of {{Interfacial Water}}},
  author = {Goertz, Matthew P. and Houston, J. E. and Zhu, X.-Y.},
  year = 2007,
  month = may,
  journal = {Langmuir},
  volume = {23},
  number = {10},
  pages = {5491--5497},
  publisher = {American Chemical Society},
  issn = {0743-7463},
  doi = {10.1021/la062299q},
  urldate = {2026-01-21}
}

@article{granickMotionsRelaxationsConfined1991,
  title = {Motions and {{Relaxations}} of {{Confined Liquids}}},
  author = {Granick, Steve},
  year = 1991,
  month = sep,
  journal = {Science},
  volume = {253},
  number = {5026},
  pages = {1374--1379},
  publisher = {American Association for the Advancement of Science},
  doi = {10.1126/science.253.5026.1374},
  urldate = {2026-01-21}
}

@article{groganBubblePatternFormation2014,
  title = {Bubble and {{Pattern Formation}} in {{Liquid Induced}} by an {{Electron Beam}}},
  author = {Grogan, Joseph M. and Schneider, Nicholas M. and Ross, Frances M. and Bau, Haim H.},
  year = 2014,
  month = jan,
  journal = {Nano Letters},
  volume = {14},
  number = {1},
  pages = {359--364},
  issn = {1530-6984, 1530-6992},
  doi = {10.1021/nl404169a},
  urldate = {2025-12-17},
  langid = {english}
}

@article{groganSituLiquidcellElectron2011,
  title = {In Situ Liquid-Cell Electron Microscopy of Colloid Aggregation and Growth Dynamics},
  author = {Grogan, Joseph M. and Rotkina, Lolita and Bau, Haim H.},
  year = 2011,
  month = jun,
  journal = {Physical Review E},
  volume = {83},
  number = {6},
  pages = {061405},
  publisher = {American Physical Society},
  doi = {10.1103/PhysRevE.83.061405},
  urldate = {2025-12-17}
}

@article{iarossiEmergenceNanofluidicsSingleBiomolecule2025,
  title = {The {{Emergence}} of {{Nanofluidics}} for {{Single-Biomolecule Manipulation}} and {{Sensing}}},
  author = {Iarossi, Marzia and Verma, Navneet Chandra and Bhattacharya, Ivy and Meller, Amit},
  year = 2025,
  month = apr,
  journal = {Analytical Chemistry},
  volume = {97},
  number = {16},
  pages = {8641--8653},
  issn = {0003-2700},
  doi = {10.1021/acs.analchem.4c06684},
  urldate = {2026-05-27},
  pmcid = {PMC12044595},
  pmid = {40244645}
}

@article{inghamXrayScatteringCharacterisation2015,
  title = {X-Ray Scattering Characterisation of Nanoparticles},
  author = {Ingham, Bridget},
  year = 2015,
  month = oct,
  journal = {Crystallography Reviews},
  volume = {21},
  number = {4},
  pages = {229--303},
  publisher = {Taylor \& Francis},
  issn = {0889-311X},
  doi = {10.1080/0889311X.2015.1024114},
  urldate = {2025-12-17}
}

@article{jamaliAnomalousNanoparticleSurface2021a,
  title = {Anomalous Nanoparticle Surface Diffusion in {{LCTEM}} Is Revealed by Deep Learning-Assisted Analysis},
  author = {Jamali, Vida and Hargus, Cory and {Ben-Moshe}, Assaf and Aghazadeh, Amirali and Ha, Hyun Dong and Mandadapu, Kranthi K. and Alivisatos, A. Paul},
  year = 2021,
  month = mar,
  journal = {Proceedings of the National Academy of Sciences},
  volume = {118},
  number = {10},
  pages = {e2017616118},
  publisher = {Proceedings of the National Academy of Sciences},
  doi = {10.1073/pnas.2017616118},
  urldate = {2025-12-17}
}

@article{jineshExperimentalEvidenceIce2008,
  title = {Experimental {{Evidence}} for {{Ice Formation}} at {{Room Temperature}}},
  author = {Jinesh, K. B. and Frenken, J. W. M.},
  year = 2008,
  month = jul,
  journal = {Physical Review Letters},
  volume = {101},
  number = {3},
  pages = {036101},
  publisher = {American Physical Society},
  doi = {10.1103/PhysRevLett.101.036101},
  urldate = {2026-01-21}
}

@article{jongeElectronMicroscopyWhole2009,
  title = {Electron Microscopy of Whole Cells in Liquid with Nanometer Resolution},
  author = {de Jonge, N. and Peckys, D. B. and Kremers, G. J. and Piston, D. W.},
  year = 2009,
  month = feb,
  journal = {Proceedings of the National Academy of Sciences},
  volume = {106},
  number = {7},
  pages = {2159--2164},
  publisher = {Proceedings of the National Academy of Sciences},
  doi = {10.1073/pnas.0809567106},
  urldate = {2026-04-13}
}

@article{kangRealspaceImagingNanoparticle2021a,
  title = {Real-Space Imaging of Nanoparticle Transport and Interaction Dynamics by Graphene Liquid Cell {{TEM}}},
  author = {Kang, Sungsu and Kim, Ji-Hyun and Lee, Minyoung and Yu, Ji Woong and Kim, Joodeok and Kang, Dohun and Baek, Hayeon and Bae, Yuna and Kim, Byung Hyo and Kang, Seulki and Shim, Sangdeok and Park, So-Jung and Lee, Won Bo and Hyeon, Taeghwan and Sung, Jaeyoung and Park, Jungwon},
  year = 2021,
  month = dec,
  journal = {Science Advances},
  volume = {7},
  number = {49},
  pages = {eabi5419},
  publisher = {American Association for the Advancement of Science},
  doi = {10.1126/sciadv.abi5419},
  urldate = {2025-12-17}
}

@article{liaoRealTimeImagingPt3Fe2012a,
  title = {Real-{{Time Imaging}} of {{Pt3Fe Nanorod Growth}} in {{Solution}}},
  author = {Liao, Hong-Gang and Cui, Likun and Whitelam, Stephen and Zheng, Haimei},
  year = 2012,
  month = may,
  journal = {Science},
  volume = {336},
  number = {6084},
  pages = {1011--1014},
  publisher = {American Association for the Advancement of Science},
  doi = {10.1126/science.1219185},
  urldate = {2026-04-13}
}

@article{liDirectionSpecificInteractionsControl2012,
  title = {Direction-{{Specific Interactions Control Crystal Growth}} by {{Oriented Attachment}}},
  author = {Li, Dongsheng and Nielsen, Michael H. and Lee, Jonathan R. I. and Frandsen, Cathrine and Banfield, Jillian F. and De Yoreo, James J.},
  year = 2012,
  month = may,
  journal = {Science},
  volume = {336},
  number = {6084},
  pages = {1014--1018},
  publisher = {American Association for the Advancement of Science},
  doi = {10.1126/science.1219643},
  urldate = {2025-12-17}
}

@article{lindforsDetectionSpectroscopyGold2004,
  title = {Detection and {{Spectroscopy}} of {{Gold Nanoparticles Using Supercontinuum White Light Confocal Microscopy}}},
  author = {Lindfors, K. and Kalkbrenner, T. and Stoller, P. and Sandoghdar, V.},
  year = 2004,
  month = jul,
  journal = {Physical Review Letters},
  volume = {93},
  number = {3},
  pages = {037401},
  issn = {0031-9007, 1079-7114},
  doi = {10.1103/PhysRevLett.93.037401},
  urldate = {2025-12-17},
  langid = {english}
}

@article{liuColloidAtomDualityAssembly2020,
  title = {``{{Colloid}}--{{Atom Duality}}'' in the {{Assembly Dynamics}} of {{Concave Gold Nanoarrows}}},
  author = {Liu, Chang and Ou, Zihao and Guo, Fucheng and Luo, Binbin and Chen, Wenxiang and Qi, Limin and Chen, Qian},
  year = 2020,
  month = jul,
  journal = {Journal of the American Chemical Society},
  volume = {142},
  number = {27},
  pages = {11669--11673},
  publisher = {American Chemical Society},
  issn = {0002-7863},
  doi = {10.1021/jacs.0c04444},
  urldate = {2025-12-17}
}

@article{liuSituVisualizationSelfAssembly2013,
  title = {In {{Situ Visualization}} of {{Self-Assembly}} of {{Charged Gold Nanoparticles}}},
  author = {Liu, Yuzi and Lin, Xiao-Min and Sun, Yugang and Rajh, Tijana},
  year = 2013,
  month = mar,
  journal = {Journal of the American Chemical Society},
  volume = {135},
  number = {10},
  pages = {3764--3767},
  publisher = {American Chemical Society},
  issn = {0002-7863},
  doi = {10.1021/ja312620e},
  urldate = {2025-12-17}
}

@article{lohMultistepNucleationNanocrystals2017,
  title = {Multistep Nucleation of Nanocrystals in Aqueous Solution},
  author = {Loh, N. Duane and Sen, Soumyo and Bosman, Michel and Tan, Shu Fen and Zhong, Jun and Nijhuis, Christian A. and Kr{\'a}l, Petr and Matsudaira, Paul and Mirsaidov, Utkur},
  year = 2017,
  month = jan,
  journal = {Nature Chemistry},
  volume = {9},
  number = {1},
  pages = {77--82},
  publisher = {Nature Publishing Group},
  issn = {1755-4349},
  doi = {10.1038/nchem.2618},
  urldate = {2026-04-13},
  copyright = {2016 Springer Nature Limited},
  langid = {english}
}

@article{luNanoparticleDynamicsNanodroplet2014,
  title = {Nanoparticle {{Dynamics}} in a {{Nanodroplet}}},
  author = {Lu, Jingyu and Aabdin, Zainul and Loh, N. Duane and Bhattacharya, Dipanjan and Mirsaidov, Utkur},
  year = 2014,
  month = apr,
  journal = {Nano Letters},
  volume = {14},
  number = {4},
  pages = {2111--2115},
  publisher = {American Chemical Society},
  issn = {1530-6984},
  doi = {10.1021/nl500766j},
  urldate = {2025-12-17}
}

@article{luoQuantifyingSelfAssemblyBehavior2017,
  title = {Quantifying the {{Self-Assembly Behavior}} of {{Anisotropic Nanoparticles Using Liquid-Phase Transmission Electron Microscopy}}},
  author = {Luo, Binbin and Smith, John W. and Ou, Zihao and Chen, Qian},
  year = 2017,
  month = may,
  journal = {Accounts of Chemical Research},
  volume = {50},
  number = {5},
  pages = {1125--1133},
  publisher = {American Chemical Society},
  issn = {0001-4842},
  doi = {10.1021/acs.accounts.7b00048},
  urldate = {2025-12-17}
}

@article{mocklSuperresolutionMicroscopySingle2020,
  title = {Super-Resolution {{Microscopy}} with {{Single Molecules}} in {{Biology}} and {{Beyond}}--{{Essentials}}, {{Current Trends}}, and {{Future Challenges}}},
  author = {M{\"o}ckl, Leonhard and Moerner, W. E.},
  year = 2020,
  month = oct,
  journal = {Journal of the American Chemical Society},
  volume = {142},
  number = {42},
  pages = {17828--17844},
  publisher = {American Chemical Society},
  issn = {0002-7863},
  doi = {10.1021/jacs.0c08178},
  urldate = {2025-12-17}
}

@article{mohantyImpermeableGraphenicEncasement2011,
  title = {Impermeable {{Graphenic Encasement}} of {{Bacteria}}},
  author = {Mohanty, Nihar and Fahrenholtz, Monica and Nagaraja, Ashvin and Boyle, Daniel and Berry, Vikas},
  year = 2011,
  month = mar,
  journal = {Nano Letters},
  volume = {11},
  number = {3},
  pages = {1270--1275},
  publisher = {American Chemical Society},
  issn = {1530-6984},
  doi = {10.1021/nl104292k},
  urldate = {2026-04-13}
}

@article{ortiz-youngInterplayApparentViscosity2013,
  title = {The Interplay between Apparent Viscosity and Wettability in Nanoconfined Water},
  author = {{Ortiz-Young}, Deborah and Chiu, Hsiang-Chih and Kim, Suenne and Vo{\"i}tchovsky, Kislon and Riedo, Elisa},
  year = 2013,
  month = sep,
  journal = {Nature Communications},
  volume = {4},
  number = {1},
  pages = {2482},
  publisher = {Nature Publishing Group},
  issn = {2041-1723},
  doi = {10.1038/ncomms3482},
  urldate = {2026-01-21},
  copyright = {2013 Springer Nature Limited},
  langid = {english}
}

@article{ouKineticPathwaysCrystallization2020,
  title = {Kinetic Pathways of Crystallization at the Nanoscale},
  author = {Ou, Zihao and Wang, Ziwei and Luo, Binbin and Luijten, Erik and Chen, Qian},
  year = 2020,
  month = apr,
  journal = {Nature Materials},
  volume = {19},
  number = {4},
  pages = {450--455},
  publisher = {Nature Publishing Group},
  issn = {1476-4660},
  doi = {10.1038/s41563-019-0514-1},
  urldate = {2025-12-17},
  copyright = {2019 The Author(s), under exclusive licence to Springer Nature Limited},
  langid = {english}
}

@article{parentTacklingChallengesDynamic2018a,
  title = {Tackling the {{Challenges}} of {{Dynamic Experiments Using Liquid-Cell Transmission Electron Microscopy}}},
  author = {Parent, Lucas R. and Bakalis, Evangelos and Proetto, Maria and Li, Yiwen and Park, Chiwoo and Zerbetto, Francesco and Gianneschi, Nathan C.},
  year = 2018,
  month = jan,
  journal = {Accounts of Chemical Research},
  volume = {51},
  number = {1},
  pages = {3--11},
  publisher = {American Chemical Society},
  issn = {0001-4842},
  doi = {10.1021/acs.accounts.7b00331},
  urldate = {2025-12-17}
}

@article{parkDirectObservationNanoparticle2012a,
  title = {Direct {{Observation}} of {{Nanoparticle Superlattice Formation}} by {{Using Liquid Cell Transmission Electron Microscopy}}},
  author = {Park, Jungwon and Zheng, Haimei and Lee, Won Chul and Geissler, Phillip L. and Rabani, Eran and Alivisatos, A. Paul},
  year = 2012,
  month = mar,
  journal = {ACS Nano},
  volume = {6},
  number = {3},
  pages = {2078--2085},
  publisher = {American Chemical Society},
  issn = {1936-0851},
  doi = {10.1021/nn203837m},
  urldate = {2025-12-17}
}

@article{pertsinidisSubnanometreSinglemoleculeLocalization2010,
  title = {Subnanometre Single-Molecule Localization, Registration and Distance Measurements},
  author = {Pertsinidis, Alexandros and Zhang, Yunxiang and Chu, Steven},
  year = 2010,
  month = jul,
  journal = {Nature},
  volume = {466},
  number = {7306},
  pages = {647--651},
  publisher = {Nature Publishing Group},
  issn = {1476-4687},
  doi = {10.1038/nature09163},
  urldate = {2025-12-17},
  copyright = {2010 Springer Nature Limited},
  langid = {english}
}

@article{powersTrackingNanoparticleDiffusion2017,
  title = {Tracking {{Nanoparticle Diffusion}} and {{Interaction}} during {{Self-Assembly}} in a {{Liquid Cell}}},
  author = {Powers, Alexander S. and Liao, Hong-Gang and Raja, Shilpa N. and Bronstein, Noah D. and Alivisatos, A. Paul and Zheng, Haimei},
  year = 2017,
  month = jan,
  journal = {Nano Letters},
  volume = {17},
  number = {1},
  pages = {15--20},
  publisher = {American Chemical Society},
  issn = {1530-6984},
  doi = {10.1021/acs.nanolett.6b02972},
  urldate = {2025-12-17}
}

@article{puseyPhaseBehaviourConcentrated1986,
  title = {Phase Behaviour of Concentrated Suspensions of Nearly Hard Colloidal Spheres},
  author = {Pusey, P. N. and {van Megen}, W.},
  year = 1986,
  month = mar,
  journal = {Nature},
  volume = {320},
  number = {6060},
  pages = {340--342},
  publisher = {Nature Publishing Group},
  issn = {1476-4687},
  doi = {10.1038/320340a0},
  urldate = {2026-04-17},
  copyright = {1986 Springer Nature Limited},
  langid = {english}
}

@article{ringVideofrequencyScanningTransmission2012,
  title = {Video-Frequency Scanning Transmission Electron Microscopy of Moving Gold Nanoparticles in Liquid},
  author = {Ring, Elisabeth A. and {de Jonge}, Niels},
  year = 2012,
  month = nov,
  journal = {Micron},
  series = {In Situ {{TEM}}},
  volume = {43},
  number = {11},
  pages = {1078--1084},
  issn = {0968-4328},
  doi = {10.1016/j.micron.2012.01.010},
  urldate = {2025-12-17}
}

@article{rossiEnvironmentalScanningElectron2004,
  title = {Environmental {{Scanning Electron Microscopy Study}} of {{Water}} in {{Carbon Nanopipes}}},
  author = {Rossi, M. P{\'i}a and Ye, Haihui and Gogotsi, Yury and Babu, Sundar and Ndungu, Patrick and Bradley, Jean-Claude},
  year = 2004,
  month = may,
  journal = {Nano Letters},
  volume = {4},
  number = {5},
  pages = {989--993},
  publisher = {American Chemical Society},
  issn = {1530-6984},
  doi = {10.1021/nl049688u},
  urldate = {2026-01-20}
}

@article{rossOpportunitiesChallengesLiquid2015,
  title = {Opportunities and Challenges in Liquid Cell Electron Microscopy},
  author = {Ross, Frances M.},
  year = 2015,
  month = dec,
  journal = {Science},
  volume = {350},
  number = {6267},
  pages = {aaa9886},
  publisher = {American Association for the Advancement of Science},
  doi = {10.1126/science.aaa9886},
  urldate = {2025-12-12}
}

@article{schneiderElectronWaterInteractions2014,
  title = {Electron--{{Water Interactions}} and {{Implications}} for {{Liquid Cell Electron Microscopy}}},
  author = {Schneider, Nicholas M. and Norton, Michael M. and Mendel, Brian J. and Grogan, Joseph M. and Ross, Frances M. and Bau, Haim H.},
  year = 2014,
  month = sep,
  journal = {The Journal of Physical Chemistry C},
  volume = {118},
  number = {38},
  pages = {22373--22382},
  publisher = {American Chemical Society},
  issn = {1932-7447},
  doi = {10.1021/jp507400n},
  urldate = {2025-12-17}
}

@article{schochTransportPhenomenaNanofluidics2008,
  title = {Transport Phenomena in Nanofluidics},
  author = {Schoch, Reto B. and Han, Jongyoon and Renaud, Philippe},
  year = 2008,
  month = jul,
  journal = {Reviews of Modern Physics},
  volume = {80},
  number = {3},
  pages = {839--883},
  publisher = {American Physical Society},
  doi = {10.1103/RevModPhys.80.839},
  urldate = {2026-02-06}
}

@article{smithElectronVideographyLipid2024,
  title = {Electron Videography of a Lipid--Protein Tango},
  author = {Smith, John W. and Carnevale, Lauren N. and Das, Aditi and Chen, Qian},
  year = 2024,
  month = apr,
  journal = {Science Advances},
  volume = {10},
  number = {16},
  pages = {eadk0217},
  publisher = {American Association for the Advancement of Science},
  doi = {10.1126/sciadv.adk0217},
  urldate = {2025-12-17}
}

\end{document}